\documentclass[preprint,aps]{revtex4}
 
\usepackage{epsf}
 
\begin{document}

\begin{center}

\title{The Coupled Cluster Method Applied to the 
{\it XXZ} Model on the Square Lattice}

\author{D.J.J. Farnell$^1$ and R.F. Bishop$^2$}
 
\affiliation{$^1$Unit of Ophthalmology, Department of Medicine,
 University of  Liverpool, Liverpool L69 3GA, United Kingdom}

\affiliation{$^2$School of Physics and Astronomy, The University of Manchester, Sackville Street Building, P.O. Box 88, Manchester, M60 1QD, United Kingdom}
                          
\date{\today}

\begin{abstract}
A review of the coupled cluster method (CCM) applied
to lattice quantum spin systems is presented here. 
The CCM formalism is explained and an application
to the spin-half {\it XXZ} model on the square 
lattice is presented. Low orders of approximation
are carried out analytically and a high-order 
CCM formulation is presented. Results for the SUB2 
approximation are carried out numerically for 
SUB2-$m$ and analytically for SUB2. It is found
that SUB2-$m$ results converge rapidly to the full 
SUB2 solution, including new results for the SUB2-$m$
limiting points compared to the SUB2 critical point. 
Results for the ground-state energy and 
the sublattice magnetisation are presented. A study
of the excitation spectrum of this model at the 
SUB2 critical point is given. Indeed, the shape 
of the excitation spectrum at the SUB2 critical point 
is identical to that predicted by spin-wave theory
for the isotropic model, albeit with a multiplicative
factor of 1.1672. This result compares very
well to results of cumulant series expansions and
Monte Carlo simulation that again predict a similar 
shape for the excitation spectrum, but with 
multiplicative factors of 1.18 and 1.21$\pm$0.03, 
respectively. Results for the isotropic Heisenberg 
model on the square lattice for the spin-one 
antiferromagnet and the spin-one/spin-half 
ferrimagnet are also given.
\end{abstract}
                                                                               
\maketitle

\end{center}

\section{Introduction}

Lattice quantum spin systems at zero temperature may be treated using a variety of methods, including exact solutions
\cite{ba1,ba2,ba3,ba4} in certain cases, as the number of lattice sites, $N$, goes to $N \rightarrow \infty$. However, the majority of cases for quantum spin systems may not be solved by using such exact techniques and so we must use approximate methods.
An advance in the treatment of low-dimensional quantum systems has been seen with the introduction of the density matrix renormalisation group (DMRG) technique
\cite{DMRG1}. This method provides  accurate results consistently for a variety of one-dimensional or quasi-one-dimensional quantum systems. However, it is still perhaps fair to say that the power of the DMRG method still remains largely untapped for systems of two or higher spatial dimensions, although we note that recent advances are encouraging
\cite{DMRG2}.
The properties of ``unfrustrated'' lattice quantum spin systems of two spatial dimensions may however be considered using the quantum Monte Carlo  (QMC) method \cite{qmc1,qmc2}. We note that QMC calculations offer the possibility of a level of accuracy limited only by the computing power available. We note that errors scale in a statistically well-understood manner with the length of the MC simulation. However, the QMC method is severely limited by the infamous sign problem, which at zero temperature is a reflection of quantum ``frustration'' in the spin system. There are many other approximate techniques which may be applied, such as spin-wave theory \cite{swt1}, exact diagonalisations \cite{ed1}, and cumulant series expansions
\cite{series1}, although these techniques are not discussed here.

We wish to focus here on another approximate technique called the coupled cluster method (CCM). The CCM often provides accurate results even in the presence of very strong frustration. Note that the CCM technique  \cite{refc1,refc2,refc3,refc4,refc5,refc6,refc7,refc8,refc9}  is a well-known method of quantum many-body theory (QMBT). We note that the CCM has been applied with much success over the last ten or so years to quantum magnetic systems at zero temperature \cite{ccm1,ccm2,ccm999,ccm3,ccm4,ccm5,ccm6,ccm7,ccm8,ccm9,ccm10,ccm11,ccm12,ccm13,ccm14,ccm15,ccm16,ccm17,ccm18,ccm19,ccm20,ccm21,ccm22,ccm23,ccm24,ccm25,ccm26}. 
In particular, the use of computer-algebraic implementations 
\cite{ccm12,ccm15,ccm20} of the CCM for quantum systems of large or infinite numbers of particles has largely been found to be very effective with respect to these spin-lattice problems.    

We present a brief description of the CCM  formalism. We then describe the application of the method to the spin-half {\it XXZ} model for the square lattice at zero temperature using two model states. The practicalities of using the CCM are illustrated by considering the LSUB2 and SUB2 approximations in detail, and their analytical solution is shown. The method of carrying out high-order CCM calculations 
using computational approaches for the localised LSUB$m$ and SUB$m$-$m$ approximation schemes 
is described in detail. A calculation for the excitation spectra of the spin-half {\it XXZ} model using the SUB2  approximation in the ground state is presented. Results for excitation energies determined using localised approximation schemes to high order is also presented. We then broaden our treatment of the Heisenberg model on the square lattice  to include quantum spin numbers of $s>1/2$. The results are found to be in good agreement with those 
results of the best of other approximate methods, where they exist.
We conclude with a discussion of the implications of these results  and of future applications of the CCM.

\section{The CCM  Formalism}

A brief description of the normal coupled cluster method (NCCM) formalism is now provided, although the interested reader is referred to Refs. \cite{refc1,refc2,refc3,refc4,refc5,refc6,refc7,refc8,refc9,ccm1,ccm2,ccm999,ccm3,ccm4,ccm5,ccm6,ccm7,ccm8,ccm9,ccm10,ccm11,ccm12,ccm13,ccm14,ccm15,ccm16,ccm17,ccm18,ccm19,ccm20,ccm21,ccm22,ccm23,ccm24,ccm25,ccm26}
for further details. 
The exact ket and bra ground-state energy eigenvectors, $|\Psi\rangle$ and $\langle\tilde{\Psi}|$, of a general many-body system described by a Hamiltonian $H$
\begin{equation} H |\Psi\rangle = E_g |\Psi\rangle\;; \;\;\;  \langle\tilde{\Psi}| H = E_g \langle\tilde{\Psi}| \;, \label{ccm_eq1} 
\end{equation} 
are parametrised within the single-reference CCM as follows:   
\begin{eqnarray} |\Psi\rangle = {\rm e}^S |\Phi\rangle \; &;&  \;\;\; S=\sum_{I \neq 0} {\cal S}_I C_I^{+}  \nonumber \; , \\ \langle\tilde{\Psi}| = \langle\Phi| \tilde{S} {\rm e}^{-S} \; &;& \;\;\; 
\tilde{S} =1 + \sum_{I \neq 0} \tilde{{\cal S}}_I C_I^{-} \; .  \label{ccm_eq2} \end{eqnarray} 
The single model or reference state $|\Phi\rangle$ is required to have the property of being a cyclic vector with respect to two well-defined Abelian subalgebras of {\it multi-configurational} creation operators $\{C_I^{+}\}$ and their Hermitian-adjoint destruction counterparts $\{ C_I^{-} \equiv (C_I^{+})^\dagger \}$. Thus, $|\Phi\rangle$ plays the role of a vacuum state with respect to a suitable set of (mutually commuting) many-body creation operators $\{C_I^{+}\}$. Note that $C_I^{-} |\Phi\rangle = 0$, $\forall ~ I \neq 0$, and that $C_0^{-} \equiv 1$, the identity operator. These operators are furthermore complete in the many-body Hilbert (or Fock) space. Also, the {\it correlation operator} $S$ is decomposed entirely in terms of these creation operators $\{C_I^{+}\}$, which, when acting on the model state ($\{C_I^{+}| 
\Phi \rangle \}$), create excitations from it. We note that although the manifest Hermiticity, ($\langle \tilde{\Psi}|^\dagger = |\Psi\rangle/\langle\Psi|\Psi\rangle$), is lost, the normalisation conditions 
$\langle \tilde{\Psi} | \Psi\rangle= \langle \Phi | \Psi\rangle = \langle \Phi | \Phi \rangle \equiv 1$ are explicitly imposed. The {\it correlation coefficients} $\{ {\cal S}_I, \tilde{{\cal S}}_I \}$ 
are regarded as being independent variables, and the full set $\{ {\cal S}_I, \tilde{{\cal S}}_I \}$ thus provides a complete description of the ground state. For instance, an arbitrary operator $A$ will have a ground-state expectation value given as, 
\begin{equation} \bar{A}\equiv \langle\tilde{\Psi}\vert A \vert\Psi\rangle=\langle\Phi | \tilde{S} {\rm e}^{-S} A {\rm e}^S | \Phi\rangle=\bar{A}\left( \{ {\cal S}_I,\tilde{{\cal S}}_I \} \right) \; . 
\label{ccm_eq6}
\end{equation} 
We note that the exponentiated form of the ground-state CCM parametrisation of Eq. (\ref{ccm_eq2}) ensures the correct counting of the {\it independent} and excited correlated many-body clusters with respect to $|\Phi\rangle$ which are present in the exact ground state $|\Psi\rangle$. It also ensures the exact incorporation of the Goldstone linked-cluster theorem, which itself guarantees the size-extensivity of all relevant extensive physical quantities. We also note that any operator in a similarity transform may be written as \begin{equation}
\tilde A 
\equiv e^{-S} A e^S 
= A + [A,S] + \frac 1{2!}[[A,S],S] + \cdots \label{ccm_nested}
\end{equation}
The determination of the correlation coefficients 
$\{ {\cal S}_I, \tilde{{\cal
 S}}_I \}$ is achieved by taking appropriate projections onto the ground-state Schr\"odinger equations of Eq. (\ref{ccm_eq1}). Equivalently, they may be determined variationally by requiring the ground-state energy expectation functional $\bar{H} ( \{ {\cal S}_I, \tilde{{\cal S}}_I\} )$, defined as in Eq. (\ref{ccm_eq6}), to be stationary with respect to variations in each of the (independent) variables of the full set. We thereby easily derive the following coupled set of equations, 
\begin{eqnarray} \delta{\bar{H}} / \delta{\tilde{{\cal S}}_I} =0 & \Rightarrow &   \langle\Phi|C_I^{-} {\rm e}^{-S} H {\rm e}^S|\Phi\rangle = 0 ,  ~ \forall ~ I \neq 0 \;\; ; \label{ccm_eq7} \\ \delta{\bar{H}} / \delta{{\cal S}_I} =0 & \Rightarrow & \langle\Phi|\tilde{S} {\rm e}^{-S} [H,C_I^{+}] {\rm e}^S|\Phi\rangle = 0 , ~ \forall ~ I \neq 0 \; . \label{ccm_eq8}
\end{eqnarray}  
Equation (\ref{ccm_eq7}) also shows that the ground-state energy at the stationary point has the simple form 
\begin{equation} 
E_g = E_g ( \{{\cal S}_I\} ) = \langle\Phi| {\rm e}^{-S} H {\rm e}^S|\Phi\rangle\;\; . \label{ccm_eq9}
\end{equation}  
It is important to realize that this (bi-)variational formulation does {\it not} lead to an upper bound for $E_g$ when the summations for $S$ and $\tilde{S}$ in Eq. (\ref{ccm_eq2}) are truncated, due to the lack of exact Hermiticity when such approximations are made. However, one can prove  that the important Hellmann-Feynman theorem {\it is} preserved in all such approximations. We note that Eq. (\ref{ccm_eq7}) represents a coupled set of non-linear multinomial equations for the  $c$-number correlation coefficients $\{ {\cal S}_I \}$. The nested commutator expansion of the similarity-transformed Hamiltonian
\begin{equation}
\tilde H \equiv e^{-S} 
H e^S = H + [H,S] + \frac
 1{2!}[[H,S],S] + \cdots 
\label{exp_H}
\end{equation} 
and the fact that all of the individual components of $S$ in the sum in Eq. ({\ref{ccm_eq2}) commute with one another, together imply that each element of $S$ in Eq. ({\ref{ccm_eq2}) is linked directly to the Hamiltonian in each of the terms in Eq. (\ref{exp_H}}). Thus, each of the coupled equations (\ref{ccm_eq7}) is of {\it linked cluster} type. Furthermore, each of these equations is of finite length when expanded, since the otherwise infinite series of Eq. (\ref{exp_H}) will always terminate at a finite order, provided only (as is usually the case) that each term in the second-quantised form of the Hamiltonian, $H$, contains a finite number of single-body destruction operators, defined with respect to the reference (vacuum) state $|\Phi\rangle$. Hence, the CCM parametrisation naturally leads to a workable scheme which can be efficiently implemented computationally. It is important to note that at the heart of the CCM lies a similarity transformation, in contrast with the unitary transformation in a standard variational formulation in which the bra state $\langle \tilde \Psi |$ is simply taken as the explicit Hermitian conjugate of $|\Psi\rangle$. In the case of spin-lattice problems of the type considered here, the operators $C_I^+$ become products of spin-raising operators $s_k^+$ over a set of sites $\{k\}$, with respect to a model state $|\Phi\rangle$ in which all spins points ``downward'' in some suitably chosen local spin axes. The CCM formalism is exact in the limit of inclusion of all possible such multi-spin cluster correlations for $S$ and $\tilde S$, although in any real application this is usually impossible to achieve. It is therefore necessary to utilise various approximation schemes within $S$ and $\tilde{S}$. The three most commonly employed schemes previously utilised have been: (1) the SUB$n$ scheme, in which all correlations involving only $n$ or fewer spins are retained, but no further restriction is made concerning their spatial separation on the lattice; (2) the SUB$n$-$m$  sub-approximation, in which all SUB$n$ correlations spanning a range of no more than $m$ adjacent lattice sites are retained; and (3) the localised LSUB$m$ scheme, in which all multi-spin correlations over all distinct locales on the lattice defined by $m$ or fewer contiguous sites are retained. The problem of solving for these types of approximation schemes using analytical and computational approaches is  discussed below.

An excited-state wave function, $|\Psi_e\rangle$, is determined by linearly applying an excitation  operator $X^e$ to the ket-state wave function of Eq. (\ref{ccm_eq2}), such that  \begin{equation}
|\Psi_e\rangle = X^e ~ e^S |\Phi\rangle ~~ .
\label{eq16}
\end{equation}
This equation may now be used to determine the low-lying excitation energies, such that the Schr{\"o}dinger equation, $H  |\Psi_e\rangle = E_e |\Psi_e\rangle$, may be combined with its ground-state counterpart of Eq. (\ref{ccm_eq1}) to give the result,
\begin{equation}
\epsilon_e X^e  | \Phi \rangle = e^{-S} [H,X^e] e^S 
| \Phi \rangle ~~ ,
\label{eq17}
\end{equation}
where $\epsilon_e \equiv E_e-E_g$ is the excitation energy. By analogy with the ground-state formalism, the excited-state correlation operator is written as,
\begin{equation}
X^e = \sum_{I \ne 0} {\cal 
X}_I^e C_{I}^+ ~~ ,
\label{eq18}
\end{equation}
where the set $\{C_{I}^+\}$ of multi-spin creation operators may differ from those used in the ground-state parametrisation in Eq. (\ref{ccm_eq2}) if the excited state has different quantum numbers than the ground state. We note that Eq. (\ref{eq18}) implies the overlap relation $\langle \Phi | \Psi_e \rangle = 0$. By applying $\langle \Phi | C_I^-$ to Eq. (\ref{eq17}) we find that,
\begin{equation}
\epsilon_e {\cal X}_I^e = \langle \Phi | C_I^- e^{-S} [H,X^e] e^S | \Phi \rangle ~~ , \forall ~ I \ne 0 ~~ , \label{temp1}
\end{equation}
which is a generalised set of eigenvalue equations with eigenvalues $\epsilon_e$ and corresponding eigenvectors ${\cal X}_I^e$, for each of the excited states which satisfy $\langle \Phi | \Psi_e \rangle = 0$. We note that lower orders of approximation may be determined analytically for both the ground and excited states. Examples of applying the LSUB2 and SUB2 approximations to the spin-half square-lattice {\it XXZ} model are given later in order to show clearly how this is performed. However, it rapidly becomes clear that analytical determination of the CCM equations for higher orders of approximation is impractical. We therefore employ computer algebraic techniques in order efficiently to determine and solve the CCM ket- and bra-state equations (discussed below). The bra-state equations may be determined easily thereafter and the ket- and bra-state equations are readily solved using standard techniques for the solution of coupled polynomial equations (e.g., the Newton-Raphson method). The excited-state eigenvalue equations may be also determined in an analogous manner, and, although this is not strictly necessary, we restrict the level of approximation to the same for the excited state as for the ground state in calculations presented here. A full exposition of the details in applying the CCM to high orders of approximation is given for the ground state in Refs. \cite{ccm7,ccm12,ccm20} and for excited states in Ref. \cite{ccm15}. Note that the results of SUB$m$-$m$ and LSUB$m$  approximation schemes may be extrapolated to the exact limit, $m \rightarrow \infty$, using various ``heuristic' approaches. How to do this is not discussed here, although the interested reader is referred to Refs. \cite{ccm12,ccm20} for more details.

\section{The {\it XXZ} Model on the Square Lattice}

We wish to apply the CCM to the spin-half {\it XXZ} model on the square lattice in order to illustrate how one applies the CCM to a practical problem. We note that this system is unfrustrated and the {\it XXZ} Hamiltonian is specified as follows,
\begin{equation}
H=\sum_{\langle i,j \rangle}
[s^x_is^x_j + s^y_is^y_j+
\Delta s^z_is^z_j]~~,
\label{xxz_1}
\end{equation}
where the sum on 
$\langle i,j\rangle$ counts 
all nearest-neighbour pairs once. The N\'eel state is the ground state in the trivial Ising limit $\Delta \rightarrow \infty$, and a phase transition occurs at (or near to) $\Delta =1$. Indeed, the ground state demonstrates N\'{e}el-like order in the $z$-direction for $\Delta >1$ and a similar $x$-$y$ planar phase for $-1<\Delta <1$. The system is ferromagnetic for $\Delta <-1$. We note that approximately 61$\%$ of the classical ordering remains in the quantum Heisenberg model at $\Delta=1$.

\subsection{The CCM applied to the {\it XXZ} model}

We turn now to the choice of model state $|\Phi\rangle$ and the operators $\{ C^+_I\}$ for the case of spin-half quantum antiferromagnets on bipartite lattices. In the regime where the corresponding classical limit is described by a N\'{e}el-like order in which all spins on each sublattice are separately aligned in some global spin axes, it is  convenient to introduce a different local quantisation axis and different spin coordinates on each sublattice. This is achieved by a suitable rotation in spin space, so that the corresponding reference state becomes a fully aligned (``ferromagnetic'') state, with all spins pointing along, say, the negative $z$-axis in the corresponding local axes. Such rotations are cannonical tranformations that leave the spin commutation relations unchanged. In the same local axes, the configuration indices $I\rightarrow\{ k_1,k_2,\cdot\cdot\cdot ,k_M\}$, a set of site indices, such that $C^+_I\rightarrow s^+_{k_{1}} s^+_{k_{2}}\cdot\cdot\cdot s^+_{k_{M}}$, where $s^{\pm}_k\equiv s^x_k \pm is^y_k$ are the usual spin-raising and spin-lowering operators at site $k$. For the Hamiltonian of Eq. (\ref{xxz_1}) we first choose the $z$-aligned N\'{e}el state as our reference state (which is the exact ground state for $\Delta \rightarrow \infty$, and is expected to be a good starting point for all $\Delta >1$, down to the expected phase transition at $\Delta =1$). We perform a rotation of the up-pointing spins by 180$^{\rm o}$ about the $y$-axis, such that $s^x\rightarrow -s^x,~s^y\rightarrow s^y,~s^z\rightarrow -s^z$ on this sublattice. The Hamiltonian of Eq. 
(\ref{xxz_1}) may thus be written in these local coordinates as,
\begin{equation}
H=-{1\over2}
\sum_{\langle i,j \rangle}
[s^+_is^+_j+s^-_is^-_j
+2\Delta s^z_is^z_j]~~.
\label{xxz_2}
\end{equation}
There is never a unique choice of model state $|\Phi\rangle$. Indeed, our choice should be guided by any physical insight available to us concerning the system or, more specifically, that particular phase of it which is under consideration. In the absence of any other insight into the quantum many-body system, we may sometimes be guided by the behaviour of the corresponding classical system. The {\it XXZ} model under consideration provides just such an illustrative example. Thus, for $\Delta >1$ the {\it classical} Hamiltonian of Eq. (\ref{xxz_1}) on the 2D square lattice (and, indeed, on any bipartite lattice) is minimized by a perfectly antiferromagnetically
N\'{e}el-ordered state in the $z$-direction, and we have already utilised this information in the preceding subsections. However, the classical ground-state energy is minimized by a N\'eel-ordered state with spins pointing along any direction in the {\it xy} plane, say along the $x$-axis for $-1<\Delta <1$. Thus, in order to provide CCM results in the region $-1<\Delta <1$, we now take this state to be our model state and we shall refer to it as the ``planar''  model state.

In order to produce another ``ferromagnetic'' model state for the planar model state in the local frames, we rotate the axes of the left-pointing spins (i.e., those pointing in the negative $x$-direction) in the planar state by $90^{\circ}$ about the $y$-axis, and the axes of the corresponding right-pointing spins by $-90^{\circ}$ about the $y$-axis. (Note that the positive $z$-axis is defined here to point upwards and the positive $x$-axis is defined to point rightwards.) Thus, the transformations of the local axes are described by \begin{equation}s^x\rightarrow s^z~~,~~~~s^y\rightarrow s^y~~,~~~~s^z\rightarrow -s^x \end{equation}for the left-pointing spins, and by \begin{equation}s^x\rightarrow -s^z~~,~~~~s^y\rightarrow s^y~~,~~~~s^z\rightarrow s^x 
\end{equation}
for the right-pointing spins. The transformed Hamiltonian of Eq. ({\ref{xxz_1}) may now be written in these local axes as \begin{equation}
H=-{\frac{1}{4}}\sum_{\langle i,j\rangle}\left[ (\Delta +1)
(s^+_is^+_j+s^-_is^-_j )
+(\Delta -1)
(s^+_is^-_j+s^-_is^+_j)+
4s^z_is^z_j\right] ~~, \label{xxz_planar_H}
\end{equation}
We note that all of the CCM correlation coefficients are zero at $\Delta=-1$ because the model state is an exact ground eigenstate of the Hamiltonian of Eq. (\ref{xxz_planar_H}) at this point. Hence, we track the CCM solution for the planar model state from this ``trivial'' point at $\Delta=-1$. 

The results presented below are based on the SUB2 approximation scheme and the localised SUB$m$-$m$ and LSUB$m$ schemes. We include all {\it fundamental configurations}, $I\rightarrow\{ k_1,k_2,\cdot\cdot\cdot k_n\}$, with $n\leq m$, which are distinct under the point and space group symmetries of both the lattice and the Hamiltonian. The numbers, $N_F$ and $N_{F_{e}}$, of such fundamental configurations for the ground and excited states, respectively, may be further restricted by the use of additional conservation laws. For example, the Hamiltonian of Eq.  (\ref{xxz_1}) commutes with the total uniform magnetisation, $s^z_T=\sum_k s^z_k$, where the sum on $k$ runs over all lattice sites. The ground state is known to lie in the $s^z_T=0$ subspace, and hence we exclude configurations with an odd number of spins or with unequal numbers of spins on the two equivalent sublattices for the $z$-aligned model state. A similar condition is imposed on clusters for the ground state based on the planar model state. Similarly for the excited states with the $z$-aligned model state, since we are only interested in the lowest-lying excitation, we restrict the choice of configurations to those with $s^z_T=\pm 1$. 

\subsection{The LSUB2 approximation for the spin-half, square-lattice  {\it XXZ} model for the $z$-aligned model state}

We start the LSUB2 calculation by specifying the commutation relations $[s_l^{\pm},s_k^z]=\mp  s_k^{\pm} \delta_{l,k}$ and $[s_l^{+},s_k^{-}]=2 s_k^{z} \delta_{l,k}$. We again note that the similarity transform may be expanded as a series of nested commutators in Eq. (\ref{ccm_nested}). We write the LSUB2 ket-state operator in the following simple form for the spin-half linear chain model, 
\begin{equation}
S = \frac {b_1}2 
\sum_i^N \sum_{\rho} 
s_i^+ s_{i+\rho}^+ ~~, 
\label{xxz_3}
\end{equation}
where $i$ runs over all sites on the square lattice and $\rho$ runs over all nearest-neighbour lattice vectors. Note that $b_1$ is the sole ket-state correlation coefficient. In this approximation we may therefore determine similarity transformed versions of the spin operators explicitly, given by
\begin{eqnarray}
\tilde s_l^+ &=& s_l^+
\nonumber \\
\tilde s_l^z &=& s_l^z + 
b_1 \sum_{\rho} 
s_l^+ s_{l+\rho}^+
\label{xxz_4} \\
\tilde s_l^- &=& 
s_l^- - 2 b_1 \sum_{\rho}  s_l^z s_{l+\rho}^+ -  
b_1^2  \sum_{\rho_1,\rho_2} s_{l}^+ s_{l+\rho_1}^+ s_{l+\rho_2}^+  
\nonumber
\end{eqnarray}
We note that the otherwise infinite series of operators in the expansion of the similarity transform terminates to finite order. We also note that $(s_l^+)^2 | \Phi \rangle =0$ for {\it any} lattice site (which is true only for spin-half systems), and this is implicitly assumed in the last of Eqs. (\ref{xxz_4}). Clearly we may also write the similarity transformed version of the Hamiltonian as
\begin{equation}
\tilde H=-{1\over 2}
\sum_{\langle i,j \rangle}
[\tilde s^+_i \tilde s^+_j
+ \tilde s^-_i \tilde s^-_j
+2\Delta \tilde s^z_i 
\tilde s^z_j]~~.
\label{xxz_5}
\end{equation}
We may now substitute the expressions for the spin operators in Eq. (\ref{xxz_4}) into the above expression. The ground-state energy is given by
\begin{equation}
\frac {E_g}N = 
- \frac 12 \{ \Delta + 2 b_1\}
~~ .
\label{xxz_6}
\end{equation}
We note that our expression for the ground-state energy is size-extensive (i.e., it scales linearly with $N$), as required by the Goldstone theorem which is obeyed by the NCCM. Furthermore, this expression terminates to finite order, as for the similarity transformed versions of spin operators. Finally, we note that {\it any} other non-trivial choice for $S$ will always yield this expression for the ground-state energy. The task is now to find $b_1$ and we note that if we could include all possible spin correlations in $S$ then we would obtain an exact result for the ground-state energy. However, this is found to be impossible to achieve for most cases in practice, and we make an approximation (such as the LSUB2 approximation presented here). The LSUB2 ket-state equation is given by
\begin{equation}
5b_1^2 + 6\Delta b_1 - 1 = 0 ~~ ,
\label{xxz_7}
\end{equation}
which therefore implies that the LSUB2 ground-state energy maybe written explicitly in terms of 
$\Delta$ as
\begin{equation}
\frac {E_g}N =  \frac {\Delta}{10} - \frac 15 \sqrt{9\Delta^2+5} ~~ .
\label{xxz_8}
\end{equation}
We note that this expression gives the correct result in the Ising limit $\Delta \rightarrow \infty$. We again note that the bra state does not manifestly have to be the Hermitian conjugate of the ket state, and we note that the bra-state correlation operator for the LSUB2 approximation is given by,
\begin{equation}
\tilde S = 1+ \frac{\tilde b_1}{2} \sum_j^N \sum_{\rho} 
s_j^- s_{j+\rho}^- ~~ ,
\label{xxz_9}
\end{equation}
where the index $j$ runs over all sites on the linear chain and $\tilde b_1$ is the sole bra-state correlation coefficient in the LSUB2 approximation. In order to determine the bra-state equation, we now explicitly determine $\bar{H}( \{ {\cal S}_I, \tilde{{\cal S}}_I\} )$, 
\begin{equation}
\bar H = -\frac N2(\Delta+2b_1) + N \tilde b_1 \Big (6\Delta b_1 + 5 b_1^2 - 1 \Big ) ~~ ,
\label{xxz_10}
\end{equation}
such that LSUB2 bra-state equation is given from ${\partial \bar H}/{\partial b_1} =0$ as
\begin{equation}
6\Delta \tilde b_1 + 10 b_1 \tilde b_1 - 1= 0 ~~ ,\label{xxz_11}
\end{equation}
which gives $\tilde b_1 = \frac 12 (9\Delta^2+5)^{-1/2}$. Finally, we note that once the values for the bra- and ket-state correlation coefficients have been determined (at a given level of approximation) then we may also obtain the values for expectation values, such as the sublattice magnetisation given by
\begin{equation}
M \equiv - \frac 1{Ns}
\langle \tilde \Psi | 
\sum_i^N s_i^z | \Psi \rangle
=  - \frac 2N  \langle \Phi |
\tilde S e^{-S} (\sum_i^N 
s_i^z) e^S | \Phi \rangle 
~~ .
\label{xxz_12}
\end{equation}
The sublattice magnetisation is written here in terms of the ``rotated'' spin coordinates. We note that this is given by,
\begin{eqnarray}
M_{{\rm LSUB2}} &=& 1 - 8 b_1 \tilde b_1 ~~ , \nonumber \\                
&=& \frac 15 \big [ 1 + \frac {12\Delta} {\sqrt{9\Delta^2+5}} \big ]
\label{xxz_13}
\end{eqnarray}
for the LSUB2 approximation.

\subsection{The SUB2 approximation for the spin-half, square-lattice {\it XXZ} model of the $z$-aligned model state}

The SUB2 approximation allows us to include all possible two-spin correlations in our wave function. We note that the SUB2 ket-state operator is given by
\begin{equation}
S = \frac12 \sum_i^N \sum_r 
b_r s_i^+ s_{i+r}^+ ~~ ,
\label{xxz_14}
\end{equation}
and that the index $i$ runs over all sites on the linear chain. Furthermore, the index $r$ runs over all lattice vectors which connect one sublattice to the other and $b_r$ is its corresponding SUB2 ket-state correlation coefficient for this vector. We again determine a similarity transformed version of the spin operators and we are able to determine the SUB2 equations, given by
\begin{equation}
\sum_{\rho} \Big \{ (1+2\Delta b_1 + 2 b_1^2) \delta_{\rho,r} - 2(\Delta+2b_1)b_r + 
\sum_{r'} b_{{r'}+\rho+\rho_1} b_{r-{r'}-\rho_1} \Big \} = 0 ~~ ,
\label{xxz_15}
\end{equation}
where $\rho$ runs over all (2D) nearest-neighbour lattice vectors and $\rho_1$ is any one of these lattice vectors. Equation (\ref{xxz_15}) may now be solved by employing a sublattice Fourier transform, given by
\begin{equation}
{\rm \Gamma}(q) = \sum_r e^{ {\rm i} r \cdot q} b_r ~~ ,\label{xxz_16}
\end{equation}
where $r$ again is a lattice vector (i.e., $r_x+r_y$ is an odd integer number for the
2D square lattice) which connects the different sublattices. This expression has an inverse given by
\begin{equation}
b_r = \frac {1}{\pi^2} \int_0^{\pi} dq_x  
\int_0^{\pi} dq_y  {\rm cos}(r_x q_x)
{\rm cos} (r_y q_y)
{\rm \Gamma}(q) ~~ .\label{xxz_17}
\end{equation}
The SUB2 equations (\ref{xxz_15}) and Eq. (\ref{xxz_16}) therefore lead to an expression for ${\rm \Gamma}(q)$ given by
\begin{equation}
{\rm \Gamma}(q) = 
\frac K { \gamma (q) } 
[ 1 \pm \sqrt{1 - k^2 
 \gamma^2 (q)} ]~~ ,
\label{xxz_18}
\end{equation}
where $K = \Delta + 2 b_1$, $k^2 = (1+2\Delta b_1 + 2 b_1^2) / K^2$, and $\gamma(q)=0.5(\rm{cos} q_x + \rm{cos} q_y)$. (Note that we choose the negative solution in Eq. (\ref{xxz_18}) in order to reproduce results in the trivial limit $\Delta \rightarrow \infty$.) These equations now yield a self-consistency requirement on the variable $b_1$ and they may be solved iteratively at a given value of $\Delta$. Indeed, we know that all correlation coefficients must tend to zero (namely, for SUB2: $b_r \rightarrow ~ 0, ~ \forall ~ r$) as  $\Delta \rightarrow \infty$ and we {\it track} this solution for large $\Delta$ by reducing $\Delta$ in small successive steps. We find that the discriminant in Eq. (\ref{xxz_18}) becomes negative at a {\it critical point}, $\Delta_c \approx 0.7985$. This is a strong indication that the CCM critical point is detecting the known quantum phase transition in the system at $\Delta=1$. Furthermore, the SUB2 approximation for the ground state may be used in conjunction with a SUB1 approximation for the excited state operator $X^e$ in Eq. (\ref{eq18})  in order to determine the excitation spectrum. We note that the excitation spectrum also becomes soft at the SUB2 critical point $\Delta_c$ (see below).  

We may also solve the SUB2-$m$ equations directly using computational techniques. Indeed, we study the limiting points of these approximations by using solution-tracking software (PITCON), which allows one to solve the SUB2-$m$ coupled non-linear equations. We again track our solution from the limit $\Delta \rightarrow \infty$ down to and beyond the limiting point and Fig. \ref{xxz_fig1} shows our results. In particular, we note that we have two distinct branches, although only the upper branch is a ``physical'' solution. We note that the CCM does not necessarily always provide an upper bound on the ground-state energy, although this is often the case for the physical solution. By tracking from a point at which we are sure of, the solution we guarantee that our solution is valid, and this approach is also used for LSUB$m$ approximations. We find that the two branches collapse onto the same line, namely, that of the full SUB2 solution, as we increase the level of SUB2-$m$ approximation with respect to $m$. We plot the positions of the SUB2-$m$ limiting points against $1/m^2$ in Fig. \ref{xxz_fig1b}, and we may see that these data points 
are found to be both highly linear and they tend to the critical value, $\Delta_c$, for the full SUB2 equations in the limit $m \rightarrow \infty$. It should be noted that the  LSUB$m$ and SUB$m$-$m$ approximations also show similar branches (namely, one ``physical'' and one ``unphysical'' branch) which appear to converge as one increases the magnitude of the truncation index, $m$. This is a strong indication that our LSUB$m$ and SUB$m$-$m$ critical points are also reflections of phase transitions in the real system and that our extrapolated LSUB$m$ and SUB$m$-$m$ results should tend to the exact solution.

\begin{figure}
\epsfxsize=9.5cm
\epsffile{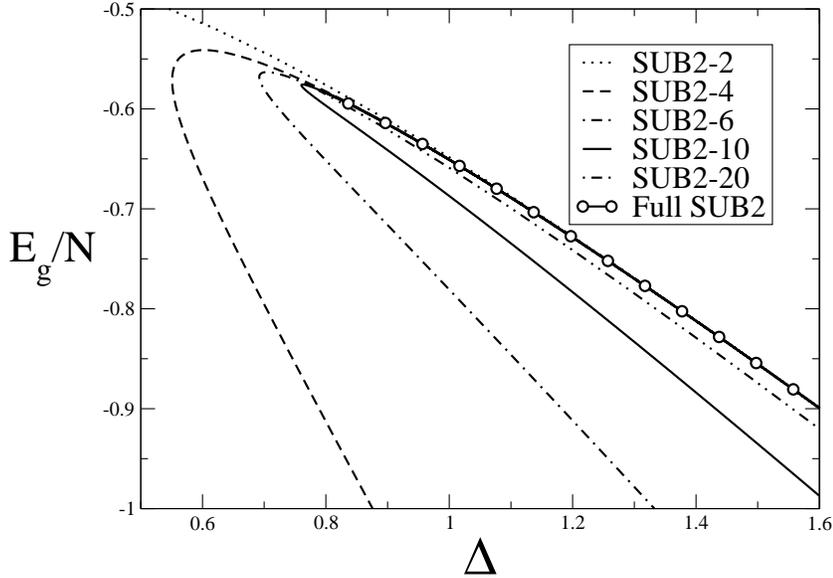}
\vspace{-5cm}
\caption{CCM SUB2-$m$ and SUB2 results using the $z$-aligned 
N\'eel model state for the ground-state energy of the spin-half
square-lattice {\it XXZ} model.}
\label{xxz_fig1}
\end{figure}

\begin{figure}
\epsfxsize=9.5cm
\epsffile{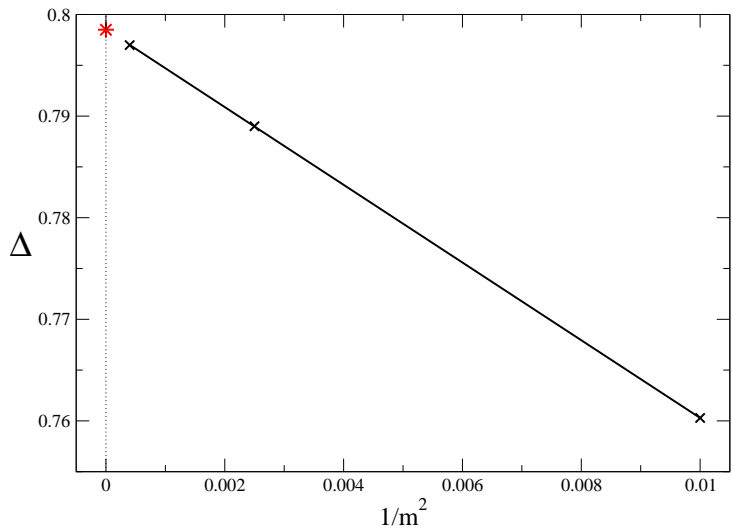}
\vspace{-5cm}
\caption{CCM SUB2-$m$ results for the limiting points 
with $m$=\{10,40,50\} for the spin-half square-lattice 
{\it XXZ} model using the $z$-aligned N\'eel model state.
The unbroken line indicates a linear fit to these three 
SUB2-$m$ limiting points, and this extrapolates to a value of 
0.79854 as $m \rightarrow \infty$. This result is in 
good agreement with the full SUB2 value for the 
critical point of 0.7985, indicated by the 
red star on the $x=0$ line.}
\label{xxz_fig1b}
\end{figure}

\subsection{High-order CCM calculations using a  computational approach}

We now consider the localised LSUB$m$ and 
SUB$m$-$m$ approximation schemes  evaluated at high orders of approximation. We note however that lower orders of approximation may be evaluated by hand, as demonstrated above. However, for higher orders of approximation we quickly find that this task becomes impossible and so we must use computational techniques in order to determine and solve the CCM equations. There are two methods of doing this. Firstly, 
we may determine the similarity transformed versions of the individual spin operators at a given level of approximation computationally and then 
substitute them into the similarity transformed version of the Hamiltonian
using computational algebra, for example. We may however have to perform further commutations of the spin operators.
This approach has the advantage of flexibility and we may consider any Hamiltonian, in principle, once the basic spin operators have been defined after similarity transformation at a given level of approximation.

Another approach is to cast the CCM ket-state correlation operator into a form given by
\begin{equation}
S = \sum_{i_1}^N 
{\cal S}_{i_1} s_{i_1}^+ +
\sum_{i_1,i_2} 
{\cal S}_{i_1,i_2} s_{i_1}^+
s_{i_2}^+ + \cdots
\label{highorder1}
\end{equation}
with respect to a model state in which all spins point in the downwards $z$-direction. Note that ${\cal S}_{i_1,\cdots,i_l}$ are the CCM ket-state correlation coefficients. We now define new operators given by 
 \begin{eqnarray}
F_k & = & \sum_l 
\sum_{i_2,\cdots,i_l}
l {\cal S}_{k,i_2,\cdots,i_l}
s_{i_2}^+ \cdots s_{i_l}^+
\nonumber \\
G_{k,m} &=& \sum_{l>1}
\sum_{i_3,\cdots,i_l} 
l(l-1) {\cal S}_{k,m,i_3,\cdots,
i_l} s_{i_3}^+ \cdots 
s_{i_l}^+
\label{highorder2}
\end{eqnarray}
for the spin-half quantum spin systems. We require additional terms for 
$s>1/2$. For example,
we may perform a similarity transform of the spin
operators, which for the spin-half system are given by
\begin{eqnarray}
\tilde s_k^+ &=& s_k^+             \nonumber \\
\tilde s_k^z &=& s_k^z + F_k s_k^+ \nonumber \\
\tilde s_k^- &=& s_k^- - 2 F_k^z - (F_k)^2 s_k^+ \label{highorder3}
\end{eqnarray} 
We now substitute these expressions into the (similarity transformed) Hamiltonian and evaluate the commutations -- but this time by hand. The Hamiltonian is then written in terms of these new operators of Eq.(\ref{highorder2}), which are themselves made up purely of spin-raising operators.

We see that this approach requires more direct effort in setting up the Hamiltonian in terms of these new operators, whereas before we have used computer algebraic techniques in order to take care of this aspect. However, once this  is accomplished, the problem of finding the ket-state equations 
then reduces to pattern matching of our 
target fundamental configurations to those terms in the Hamiltonian. 
We note that this form is then perfectly suited to a computational implementation because no further commutations
or re-ordering of terms in the Hamiltonian is necessary. 
We note that the bra-state equations may be directly determined once the ground-state energy and CCM ket-state equations have been determined \cite{ccm12,ccm15,ccm20}. 

Results for the ground-state energy of the spin-half square-lattice {\it XXZ} model are shown in Fig. \ref{xxz_fig2} and for the spin-half Heisenberg model ($\Delta=1$) in Table \ref{xxz_tab1}. We note that good correspondence with the results of the best and most accurate of other approximate methods is observed. We see clearly from Fig. \ref{xxz_fig2} that the results based on the $z$-aligned model give the best results in the region $\Delta > 1$, and those based on the planar model state appear to work best in the region $-1 < \Delta < 1$. This is in agreement with our understanding of this system  that states that we have N\'eel ordering in the $xy$-plane for $-1 < \Delta < 1$ and N\'eel ordering in the $z$-direction for $\Delta > 1$. We note that the results for the two model states are identical at $\Delta=1$, as we expect.

\begin{figure}
\epsfxsize=9.5cm
\epsffile{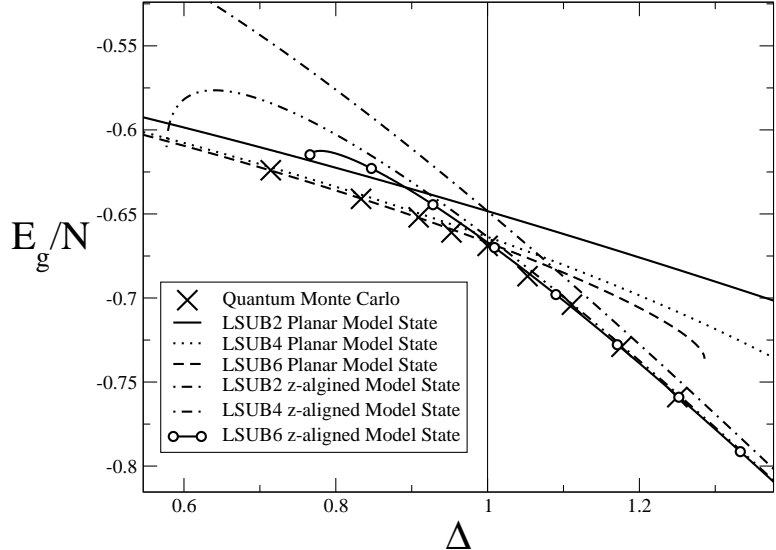}
\vspace{-5cm}
\caption{CCM LSUB$m$ results using the $z$-aligned and planar
N\'eel model states for the ground-state energy of the spin-half
square-lattice {\it XXZ} model compared to quantum Monte Carlo results
of Ref. \cite{qmc1}. Results for the LSUB6 approximation using both model
states end at their respective {\it critical} points.}
\label{xxz_fig2}
\end{figure}

\begin{figure}
\epsfxsize=9cm
\epsffile{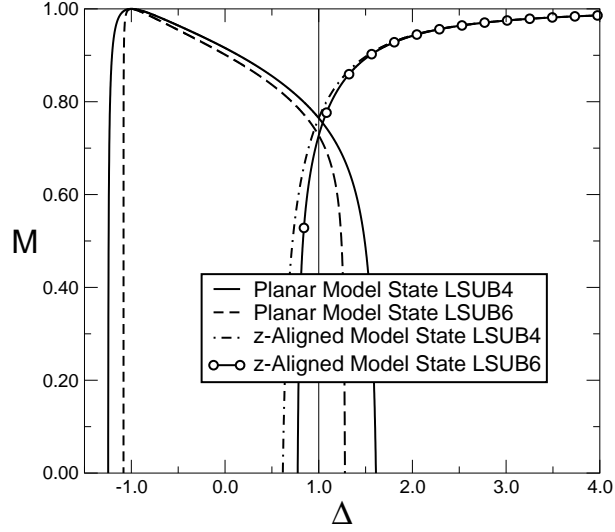}
\vspace{-5cm}
\caption{CCM LSUB$m$ results using the $z$-aligned and planar
N\'eel model states for the sublattice magnetisation of the spin-half
square-lattice {\it XXZ} model.}
\label{xxz_fig3}
\end{figure}

\subsection{Excitation Spectra of the spin-half 
square-lattice {\it XXZ} Model for the $z$-aligned model state}

We now consider the excitation spectrum, which
may be determined using the CCM for the SUB2 approximation in the ground-state by assuming
\begin{equation}
X =  \sum_i a_i s^+_i
~~.
\label{exc1}
\end{equation}
We note that the substitution of the expressions in Eqs. (\ref{xxz_14}) and (\ref{exc1}) for the ground- and excited-state operators, respectively, leads to an expression for the excited-state correlation coefficients, given by 
\begin{equation}
\frac 12 z K a_l - \frac 12
\sum_{\rho,r} b_r a_
{l+r+\rho} = 
\epsilon_e a_l
\label{exc2}
\end{equation}
(We have $z=4$ here.) This equation may be solved by Fourier transform techniques such that the an expression  for the excitation spectra is given by
\begin{equation}
\epsilon(q) = \frac 12 z K
\sqrt{1-k^2 \gamma^2(q)}
\label{exc3}
\end{equation}
We note that $K = \Delta + 2 b_1$, $k^2 = (1+2\Delta b_1 + 2 b_1^2) / K^2$, and $\gamma(q)=0.5(\rm{cos} q_x + \rm{cos} q_y)$, and that we solve the SUB2 ket-state equations 
(for a given value of the anisotropy parameter, $\Delta$, here) in order to obtain a value for $b_1$. This is substituted into Eq. (\ref{exc3}) above and so we obtain values for the excitation spectra as a function of the wave vector. We note that the excitation spectrum becomes soft at the CCM SUB2 critical point and results for the spectrum are presented in Fig. \ref{xxz_fig4}. We see that 
the CCM excitation spectrum is identical in shape to those results of 
linear spin-wave theory (SWT) with a multiplicative factor of 1.1672. 
This agrees well with results of series expansions \cite{spectra1} and 
quantum Monte Carlo \cite{spectra2} that also predict a curve identical 
to SWT with multiplicative factors 1.18 and 1.21$\pm$0.03, respectively. 

\begin{figure}
\centerline{\epsfxsize=7.5cm\epsffile{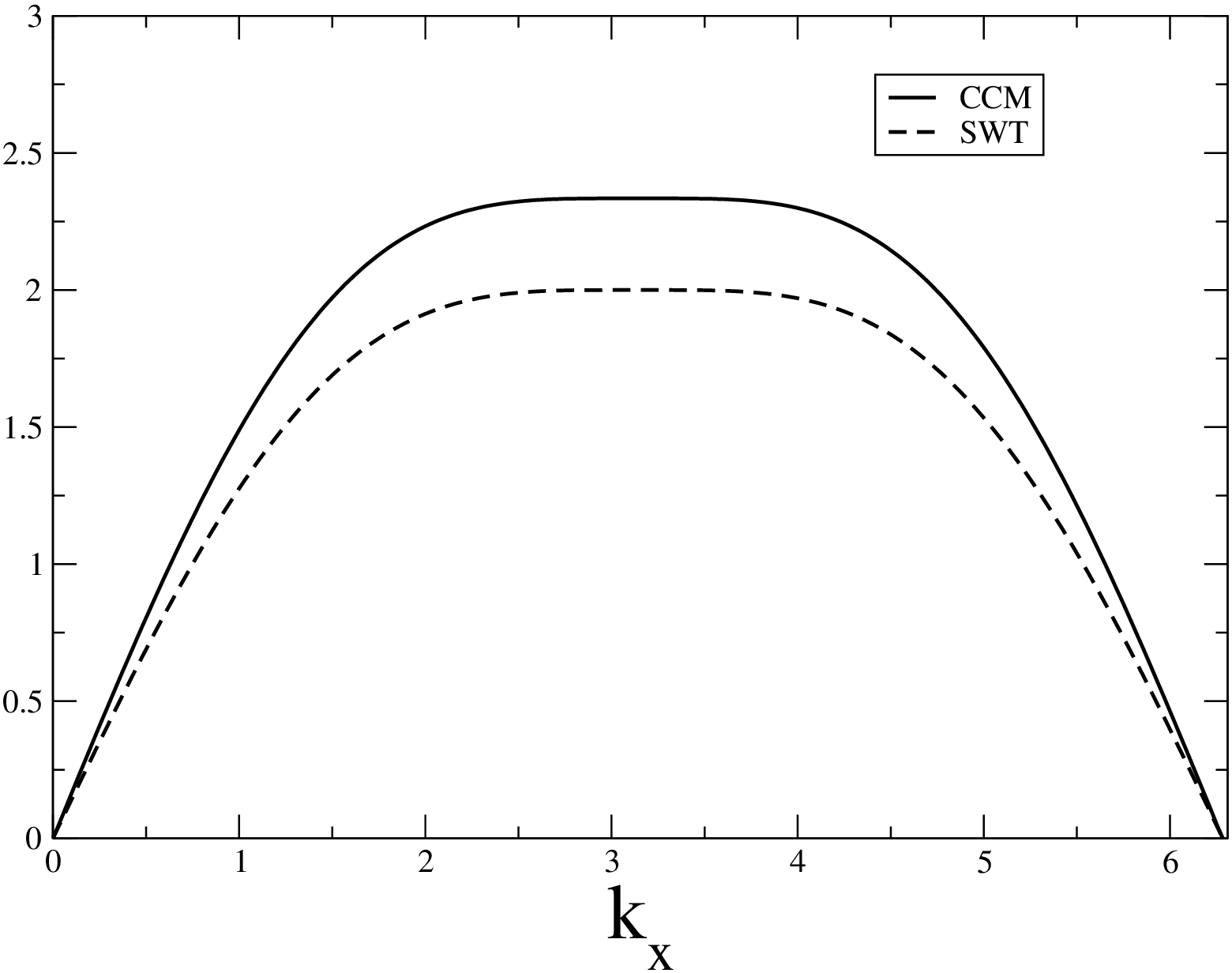} 
            \hspace*{1.0cm}
            \epsfxsize=7.5cm\epsffile{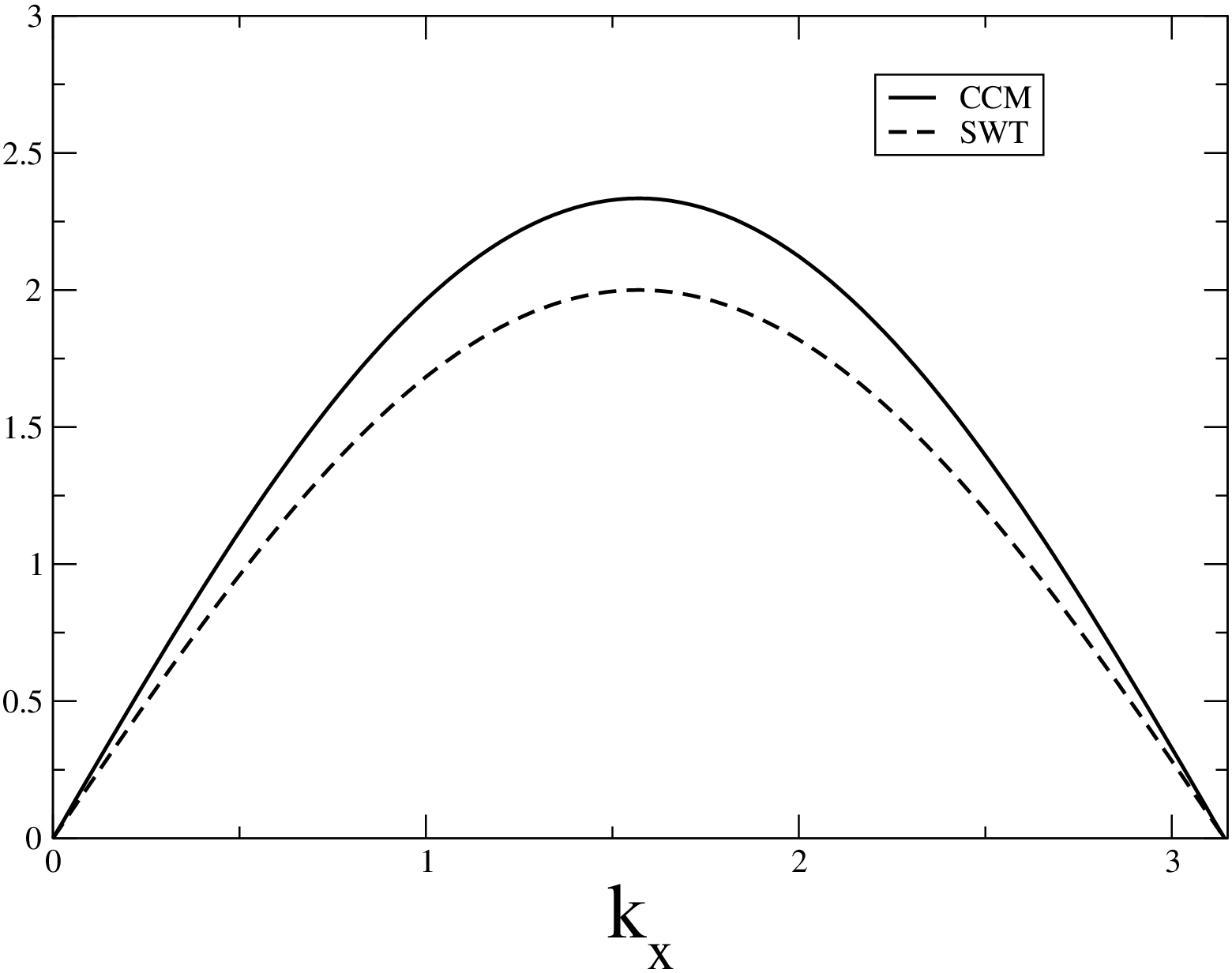}}
\vspace{-4cm}
\caption{Excitation spectrum for the Heisenberg model determined at the  
critical point at $\Delta_c=0.7985$ for the CCM results and at $\Delta=1$ 
for the spin-wave theory results. The diagram plotted on the left is for
$k_x=k_y$ and the diagram on the right is for $k_y=0$.
Note that the CCM excitation spectrum
is identical in shape to those results of SWT with a multiplicative 
factor of 1.1672. This agrees well with results of series expansions 
\cite{spectra1} and quantum Monte Carlo \cite{spectra2} 
that also predict a curve identical to SWT with multiplicative 
factors 1.18 and 1.21$\pm$0.03, respectively. Our results are in thus in 
good agreement with the best of other methods and this is further evidence
that the CCM SUB2 critical point is an indication of the quantum phase
transition at $\Delta=1$ in the ``real'' system.}
\label{xxz_fig4}
\end{figure}

Finally, we again note that the excitation energy may be determined directly from Eq. (\ref{temp1}) in ``real space'' without recourse to Fourier transform methods,
although computational techniques are again necessary for higher orders of approximation. For the sake of consistency, we often retain the same level of localised approximation 
for the ground and excited states. For example, high-order CCM results are presented for the {\it XXZ} model in Fig. \ref{xxz_fig5} and for the  Heisenberg model in Table 
\ref{xxz_tab1}. We see that the CCM results converge rapidly with LSUB$m$ approximation level. Indeed, extrapolated results predict that the excitation is gapless at $\Delta=1$, as is believed to occur for the Heisenberg model from the results of other approximate calculations (as discussed above).

\begin{figure}
\epsfxsize=9.5cm
\epsffile{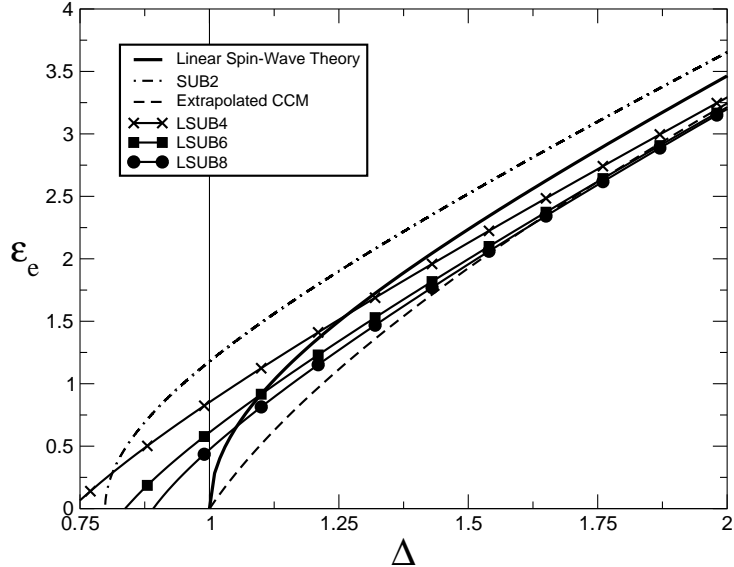}
\vspace{-5cm}
\caption{CCM LSUB$m$ results using the $z$-aligned 
N\'eel model state for the excited-state energy of the spin-half
square-lattice {\it XXZ} model compared to linear spin-wave theory
of Ref. \cite{qmc1}. }
\label{xxz_fig5}
\end{figure}

\begin{table}[t]
\caption{CCM results \cite{ccm12,ccm15} for the isotropic ($\Delta=1$) spin-half square-lattice Heisenberg antiferromagnet compared to results of other methods. The numbers of fundamental configurations in the ground-state and excited-state CCM wave functions for the $z$-aligned N\'eel model state are given by $N_f^z$ and $N_{f_e}^z$, respectively, and the number of fundamental configurations in the ground-state CCM wave function for the planar N\'eel model state is given by $N_f^p$. Results for the critical points of the $z$-aligned N\'eel model state are indicated by $\Delta_c^z$ and results for the critical points of the planar N\'eel model state are indicated by $\Delta_c^p$. (Note that results for the ground-state expectation values for both model states are identical for the isotropic Heisenberg model at $\Delta=1$.)}
\begin{center}
{\footnotesize
\begin{tabular}{|l|c|c|c|c|c|c|c|c|}  \hline\hline
Method  &$E_g/N$        &$M$            &$\epsilon_e$    
        &$N_f^z$        &$N_{f_e}^z$    &$N_f^p$
        &$\Delta_c^z$   &$\Delta_c^p$           \\ \hline\hline
LSUB2   &$-$0.64833     &0.841          &1.407           
        &1              &1              &1
        &--             &--     \\ \hline
SUB2    &$-$0.65083     &0.827          &1.178           
        &--             &--             &--
        &0.7985          &1.204     \\ \hline
LSUB4   &$-$0.66366     &0.765          &0.852           
        &7              &6              &10
        &0.577          &1.648  \\ \hline
LSUB6   &$-$0.66700     &0.727          &0.610           
        &75             &91             &131
        &0.763          &1.286  \\ \hline
LSUB8   &$-$0.66817     &0.705          &0.473           
        &1273           &2011           &2793
        &0.843          &--     \\ \hline
Extrapolated CCM 
        &$-$0.6697      &0.62   	&0.00            
        &--             &--             &--
        &1.03           &--     \\ \hline
LSWT\cite{swt1}         
        &$-$0.658       &0.606          &0.0             
        &--             &--             &--
        &1.0            &--     \\ \hline
Series Expansions\cite{series1}   
        &$-$0.6693(1)  &0.614(2)       &--              
        &--             &--             &--
        &--             &--     \\ \hline
QMC\cite{qmc2}
        &$-$0.669437(5) &0.6140(6)      &--  
        &--             &--             &--
        &--             &--     \\ \hline
\end{tabular}
}
\end{center}
\label{xxz_tab1}
\end{table}

\subsection{$s > 1/2$ Results}

We note that results for the case $s>1/2$ may be generated using 
the CCM in the same way as for $s=1/2$ for N\'eel-like model 
states. However, in this case we relax the condition that 
$(s^+)^2 | \Phi \rangle=0$, which is true for the spin-half system. 
We are now able to treat the  LSUB$m$, SUB$m$-$m$, and SUB2 
approximations in directly equivalent ways as to the 
spin-half case. We note however that new terms are 
generated  for the high-order formalism and that this 
must be taken into account for $s>1/2$ systems. How
this is achieved is explained in Ref. \cite{ccm20}.
Results for the ground-state properties of the spin-one 
antiferromagnet on the square-lattice and the spin-half/spin-one
(anti-)ferrimagnet are presented in Tables \ref{xxz_tab2}
and \ref{xxz_tab3}, respectively. 
We see that good results are obtained in comparison to other 
approximate methods. Results using the CCM, spin-wave theory, and 
exact diagonalisations have been found to be very useful in 
determining the phase diagrams of the spin-half/spin-one
ferrimagnet, even in the presence of strong frustration
\cite{ccm21}. Clearly, there are many other quantum
magnetic systems that can be treated in a similar fashion.

\begin{table}
\caption{CCM results for the ground state of the spin-one 
Heisenberg antiferromagnet at $\Delta =1$ on the square lattice using 
the LSUB$m$ approximation with $m=\{2,4,6\}$. Values for the 
CCM critical points, $\Delta_c$, of the anisotropic model 
as a function of the anisotropy, $\Delta$, are also presented.
Note that $N_F$ indicates the number of fundamental clusters
at each level of approximation.}
\begin{center}
\begin{tabular}{|@{~}c@{~}|@{~}c@{~}|@{~}c@{~}|@{~}c@{~}|@{~}c@{~}|}  \hline 
            &$N_F$  &$E_g/N$       &$M$       &$\Delta_c$    \\ \hline\hline
LSUB2       &2      &$-$2.295322   &0.909747  &0.3240        \\ \hline
SUB2\footnote{ See Ref.  \cite{ccm2}}        
            &--     &$-$2.302148   &0.8961    &0.9109        \\ \hline
LSUB4       &30     &$-$2.320278   &0.869875  &0.7867        \\ \hline
LSUB6       &1001   &$-$2.325196   &0.851007  &0.8899        \\ \hline
LSUB$\infty$ &--    &$-$2.3292     &0.8049    &0.98         \\ \hline
SWT\footnote{ See Ref.  \cite{swt1}}          
            &--     &$-$2.3282     &0.8043    &--            \\ \hline
Series Expansions\footnote{ See Ref.  \cite{series1}}
            &--     &$-$2.3279(2)  &0.8039(4) &--            \\ \hline
\end{tabular}
\end{center}
\label{xxz_tab2}
\end{table}

\begin{table}
\caption{Results for the ground-state energy and amounts
of sublattice magnetizations $M_A$ and $M_B$, on the
spin-one and spin-half sites respectively, of the 
square-lattice spin-half/spin-one $J_1$--$J_2$ ferrimagnet at $J_1=1$ and 
$J_2=0$ based on the N\'eel model state. Note that $N_F$ indicates
the number of fundamental configurations at a given 
level of  LSUB$m$ or SUB$m$-$m$ approximation. 
CCM results are compared to exact diagonalisations of 
finite-sized lattices. Heuristic 
extrapolations in the limit 
$m \rightarrow \infty$ are performed and a rough estimate 
of the error (in the last significant figure shown) is also 
given.}
\begin{center}
\begin{tabular}{|c|c|c|c|c|}  \hline 
        &$N_F$
	&$E_g/N$
        &$M_A$
        &$M_B$ \\ \hline 
SUB2-2  &1
	&$-$1.192582 
        &0.92848        
	&0.85695 \\ \hline 
SUB4-4  &13
	&$-$1.204922 
        &0.90781        
        &0.81562   \\ \hline      
SUB6-6  &268
	&$-$1.206271
        &0.90333        
	&0.80667\\ \hline 
LSUB6   &279   
	&$-$1.206281 
        &0.90329
	&0.80659\\ \hline 
Extrapolated CCM    
	&--   
	&$-$1.2069(2)
        &0.898(1)
	&0.796(2)\\ \hline 
\end{tabular} 
\end{center}
\label{xxz_tab3}
\end{table}

\section{Conclusions}

The underlying formalism and practical application of the CCM have been discussed here. 
A detailed exposition of the application of the CCM to the spin-half square-lattice {\it XXZ} model was given, including details of analytical LSUB2 and SUB2 calculations. The details of applying the CCM to high orders of approximation for a localised approximation scheme using a computational approach were explored. 
Our results were seen to be in excellent agreement with the best of other approximate methods.
The application of the CCM using different model states and for $s>1/2$ systems was also described. It was seen that excellent results can be obtained for such cases. We note that we may employ the symmetries of the lattice and Hamiltonian in order to reduce the complexity of our problem -- for example, by reducing the number of fundamental configurations at a given level of approximation.

The application of the CCM using non-N\'eel model states is an interesting possibility in the future. 
Such model states might be dimer-solid or plaquette-solid model states, for example.
This would have the advantage that novel ordering of quantum spin systems could be considered directly via the CCM. We would expect to obtain similar accuracy as that seen here using our high-order computational techniques, and this might be further extended by using parallel processing. It has recently been proven \cite{ccm22,ccm24} that numbers of fundamental clusters approaching $10^6$ are now possible, in principle. We might also wish to use  high-order CCM using computational techniques for non-N\'eel model states in order to treat the excitation spectra and excitation energies. This would provide a powerful viewpoint into such novel-ordered states in cases where other methods might typically fail, e.g., due to strong frustration. Lattice quantum spin systems might provide a possible useful arena in which to test the properties of the extended coupled cluster method (ECCM) \cite{ccm14} due to their underlying simplicity of their formalism and yet their complexity of behaviour for large numbers of particles.
 It would also be interesting to apply the CCM at non-zero temperatures, and, again, spin systems are an excellent candidate for such treatment. It would be interesting to see if some of the insight gained into quantum phase transitions at zero-temperature using the CCM might also be seen in such non-zero temperature CCM calculations.

\section{Acknowledgments}

The authors gratefully acknowledge the help and advice of John B. Parkinson,
Johannes Richter, Chen Zeng, Joerg Schulenburg, Sven Kr\"uger, Rachid Darradi,
and Roger Hale in carrying out these calculations.

\end{document}